\documentclass[aps,twocolumn,amsmath,amssymb,showpacs]{revtex4}

\usepackage{graphicx}
\usepackage{dcolumn}
\usepackage{bm}

\begin{document}

\title{Rossby wave equilibria and zonal jets}

\author{Peter B. Weichman}

\affiliation{BAE Systems, Advanced Information Technologies, 6 New
England Executive Place, Burlington, MA 01803}

%


\date{\today}

\begin{abstract}

The problem of coherent vortex and zonal jet formation in a system
of nonlinear Rossby waves is considered from the point of view of
the late time steady state achieved by free decay of a given
initial state. Statistical equilibrium equations respecting all
conservation laws are constructed, generalizing those derived
previously for 2D inviscid Euler flow.  Jet-like solutions are
ubiquitous, with large coherent vortices existing only when there
is uniform background flow with the precise velocity to cancel the
beta effect.

\end{abstract}

\maketitle



The formation of jets (localized, elongated energetic flows) in
rotating 2D flow, such as planetary atmospheres and oceans, is a
ubiquitous phenomenon. Although jet-like structures are often
dictated by external forcing, or by boundary constraints, zonal
(east-west) jets may also form via unforced evolution of
essentially isotropic initial conditions in large systems where
boundaries are unimportant \cite{Rhines}---the remarkable band
structure of Jupiter is a famous example. Although the planetary
rotation axis clearly distinguishes the zonal and meridional
directions, its effect on the dynamics is rather subtle and there
is no simple argument why this should lead to highly elongated
structures.  In particular, conserved quantities like energy $E$
and enstrophy $\Omega_2$, as well nonlinear advection effects,
remain isotropic.

Recently, a new, highly anistropic adiabatic invariant $B$,
special to systems of interacting Rossby waves \cite{B91}, has
been argued to provide the required dynamical mechanism
\cite{B05}. In particular, for the ever larger scale flows
generated by the well-known inverse cascade of energy, $B$
increases strongly with scale if the energy spectrum remains
isotropic.  Conservation of $B$ requires that the spectrum
concentrate in the meridional direction, leading to zonally
concentrated flows.  Well defined Rossby waves exist only at
mid-latitudes, so this offers only a partial explanation since jet
formation is also seen in near-equatorial flows.

In this work a very general, complementary approach is considered.
Rather than trying to follow the very complicated turbulent
dynamics from some given initial condition, one seeks to
understand only the range of possible long-time steady states
produced by free decay of such flows.  In particular,
\emph{equilibrium} steady states are considered, which are
completely specified by the values of all the conserved
quantities. In standard 3D thermodynamic systems, often only
energy and particle number are conserved. In the 2D flows of
interest, there are an infinite number, providing a far greater
range of interesting equilibria with nontrivial spatial structure
\cite{Miller,MWC,WP,CC}.

In the following, an exact nonlinear PDE that determines the
equilibrium flows is derived, and solved in various limits.  It
will be shown, quite generally, that the latitude dependence of
the Coriolis force (the ``beta effect'') destabilizes large
coherent vortex structures: the only stable equilibrium flows are
those that are zonally translation invariant, i.e., jet-like. Only
in a background flow that precisely cancels the beta effect are
coherent vortices stable. Interestingly, conservation of $B$,
which requires a beta effect, also fails in this limit, and
therefore does not hinder the formation of such structures. The
equilibrium arguments are independent of all dynamical
considerations, and hence do not depend on the existence of $B$,
but the fact that they are so consistent suggests a relation. This
will be a subject of future investigation.

For generality, consider a class of equations (examples to follow)
which may be written in the form
\begin{equation}
\partial_t Q + {\bf v} \cdot \nabla Q = 0.
\label{1}
\end{equation}
The incompressible velocity field ${\bf v} = \nabla \times \psi
\equiv (\partial_y \psi,-\partial_x \psi)$.  The stream function
$\psi$ is assumed related to the convectively conserved ``charge
field'' $Q$ through an energy functional ${\cal H}[Q]$:
\begin{equation}
\psi({\bf r}) = \delta {\cal H}/\delta Q({\bf r}).
\label{2}
\end{equation}
An example is the Charny-Hasagawa-Mima (CHM) equation in which $Q
= \omega + k_R^2 \psi + f$, where $\omega = \nabla \times {\bf v}
\equiv \partial_x v_y - \partial_y v_x = -\nabla^2 \psi$ is the
vorticity. The Coriolis function $f = 2 \Omega_E \sin[\theta(y)]$,
where $\theta(y)$ is the latitude, and $\Omega_E = 2\pi/(24$ hr)
is the rotation frequency of the Earth, and coordinates are chosen
so that the $x$-axis points eastward and the $y$-axis northward.
In this case,
\begin{equation}
{\cal H} = \frac{1}{2} \int d^2r \int d^2r'
[Q({\bf r})-f({\bf r})] g({\bf r},{\bf r}')
[Q({\bf r}')-f({\bf r}')],
\label{3}
\end{equation}
in which $g$ is the Green function of the operator $-\nabla^2 +
k_R^2$ satisfying free slip and/or periodic boundary conditions.
In the beta-plane approximation, $f = f_0 + \beta_f y$, where $f_0
= 2\Omega_E \sin(\theta_0)$, $\beta_f = 2(\Omega_E/R_E)
\cos(\theta_0)$, where $R_E$ is Earth's radius, and $\theta_0$ is
a reference latitude.  The Rossby radius of deformation is $R_0 =
1/k_R = c/f$, where $c \sim 3$ m/s is the speed of internal
gravity waves. In principle $k_R = k_R(y)$, but usually one sets
$k_R = 1/R_0(0)$. Only $\beta_f$ then produces anisotropy. The
free space Green function is the modified Bessel function, $g({\bf
r},{\bf r}') = K_0(k_R|{\bf r}-{\bf r}'|)/2\pi$, and linearizing
(\ref{1}) produces the usual Rossby wave dispersion relation
$\omega = -\beta_f k_x/(k_R^2 + k^2)$. The CHM equation (reducing
to the Euler equation for $k_R = 0$) is an approximation to the
shallow water equations in which propagating gravity waves are
neglected. The surface height is then proportional to $\psi$, and
adiabatically follows the flow. This equation also describes drift
wave plasmas, in the 2D plane perpendicular to an applied magnetic
field, where $\psi$ is the electrostatic potential and $\omega$
the charge density.

For later purposes, it is useful to define the Legendre transform
of ${\cal H}$,
\begin{equation}
{\cal L}[\psi] = {\cal H}[Q]
- \int d^2r \psi({\bf r}) Q({\bf r}),
\label{4}
\end{equation}
in which (\ref{2}) is used to substitute $\psi$ for $Q$.  The
relation may be inverted via the implied identity
\begin{equation}
Q({\bf r}) = -\delta {\cal L}/\delta \psi({\bf r}).
\label{5}
\end{equation}
From (\ref{3}) one obtains the simple result
\begin{equation}
{\cal L} = -\int d^2r \left[\frac{1}{2} |\nabla \psi|^2
+ \frac{1}{2} k_R^2 \psi^2 + f \psi \right],
\label{6}
\end{equation}
the first two terms of which may be recognized as the kinetic and
potential energies.

Since ${\bf v} \cdot \nabla \psi \equiv  0$, it follows that
${\cal H}$ is conserved \cite{foot:poisson}.  With periodic or
free slip boundary conditions, it follows also from (\ref{1}) that
$\Omega_h = \int d^2r h[Q({\bf r})]$ is conserved for any 1D
function $h(\sigma)$, conveniently summarized by
\begin{equation}
g(\sigma) = \int d^2r \delta[\sigma - Q({\bf r})],
\label{7}
\end{equation}
conserved for any $\sigma$. One recovers $\Omega_h = \int d\sigma
h(\sigma) g(\sigma)$.  Certain ``momentum functionals''
\begin{equation}
P = \int d^2 r \lambda({\bf r}) Q({\bf r})
\label{8}
\end{equation}
are also conserved if ${\cal H}$ has appropriate translation
symmetries \cite{foot:sym}. For rotational symmetry, the conserved
(vertical component of) angular momentum corresponds to $\lambda =
\frac{1}{2} r^2$. For translation symmetry along direction ${\bf
\hat l}$, the conserved linear momentum corresponds to $\lambda =
{\bf \hat l} \times {\bf r}$.

For the CHM equation in the open beta-plane there is an
additional \emph{adiabatically} conserved quantity, which to
quadratic order in $\psi$ takes the Fourier space form \cite{B91}
\begin{equation}
B = \frac{1}{2} \int \frac{d^2k}{(2\pi)^2}
\hat b({\bf k}) \hat {\cal Q}(-{\bf k}) \hat {\cal Q}({\bf k}),
\label{9}
\end{equation}
in which $\hat {\cal Q}({\bf k}) = (k^2+k_R^2) \hat \psi({\bf k})$
is the Fourier transform of $Q - f = (-\nabla^2 + k_R^2) \psi$.
The key properties are \cite{B05}: (1) $\hat b({\bf k}) = {\cal
O}[(k_R/k)^6]$ decays rapidly for $k/k_R \gg 1$, and (2) for
$k/k_R \ll 1$, $\hat b({\bf k}) = {\cal O}(1)$ for ${\bf k}$-space
directions $\pi/3 < \hat \theta < 2\pi/3$ (but vanishing for
$|\hat \theta| \to \pi/2$) while $\hat b({\bf k}) = {\cal
O}(k_R/k)$ otherwise. Property (1) implies that $B$, even more so
than ${\cal H}$, is a large scale quantity, insensitive to small
scale variations in $Q$. Property (2) then implies that the
inverse cascade must focus the small $k$ part of the spectrum
close to $|\hat \theta| = \pi/2$, leading (through the curl
relation with ${\bf v}$) to large scale zonal flows \cite{B05}.

When $k_R \to 0$ or $\beta_f \to 0$, the adiabatic conditions
fail, and $B$ is no longer conserved \cite{B91}, so this form
should be used only at mid-latitudes. Since $B$ is not exactly
conserved, it will not be included explicitly in the equilibrium
calculation, but its consistency with various proposed equilibrium
states will be confirmed at the end.

All statistical equilibrium information is contained in the free
energy
\begin{equation}
{\cal F} = -\frac{1}{\beta} \ln\left\{\int DQ \Delta_g[Q]
e^{-\beta ({\cal H}[Q] - \alpha_0 P[Q])} \right\},
\label{10}
\end{equation}
in which $\beta = 1/T$ is the inverse temperature, $\alpha_0$ is a
Lagrange multiplier for $P$, $\int DQ$ is a functional integral
over all possible configurations of $Q$, and $\Delta_g[Q] =
\prod_\sigma \delta(g(\sigma)-g_\sigma[Q])$ represents the
infinite product of delta-functions required to impose the chosen
values $g(\sigma)$ of all the conserved integrals $g_\sigma[Q]$
represented by the right hand side of (\ref{7}). If $Q({\bf r})
\to Q_i$ is discretized on a grid with microscopic spacing $a$,
then Liouville's theorem specifies the measure $\int DQ =
\lim_{a\to 0} \prod_i \int_{-\infty}^\infty dQ_i$ \cite{MWC}.

It transpires that ${\cal F}$ may be computed exactly under the
very general assumption that ${\cal H}$ and $P$ are insensitive to
very small scale fluctuations in $Q$ \cite{Miller,MWC}. Due to
fine-scale mixing, the equilibrium $Q$ will fluctuate wildly from
grid point to grid point, but its average $Q_0({\bf r})$ over a
small area $l^2 \gg a^2$ (with $l \to 0$ also at the end), will
vary smoothly. If $n_0({\bf r},\sigma)$ is the equilibrium
probability density for finding a parcel of fluid with charge
$\sigma$ in the area $l^2$ about ${\bf r}$. Then $Q_0({\bf r}) =
\int \sigma d\sigma n_0({\bf r},\sigma)$. It is assumed that
${\cal H}$ is smooth on scale $l$, and hence ${\cal H}[Q] = {\cal
H}[Q_0]$. In (\ref{3}) this is provided by the smoothness of $g$:
although it diverges at the origin, the logarithmic singularity is
sufficiently weak that this assumption remains valid \cite{MWC}.
By integrating out the small-scale fluctuations, which may be
treated as independent from grid point to grid point, one obtains
an entropy contribution,
\begin{equation}
S = \frac{1}{a^2} \int d^2r d\sigma
n_0({\bf r},\sigma) \ln [n_0({\bf r},\sigma)],
\label{11}
\end{equation}
in terms of which the free energy is ${\cal F}[n_0] = {\cal
H}[Q_0] - T S[n_0]$. One now observes that if $T S$ is to remain
finite as $a \to 0$ (so that nontrivial equilibria are obtained in
which entropy and energy compete), one must adopt the scaling $T =
\bar T a^2$, $\beta = \bar \beta/a^2$, with fixed $\bar \beta =
1/\bar T$.

To self consistently determine $n_0({\bf r},\sigma)$, Lagrange
multipliers are introduced via the Gibbs free energy
\begin{equation}
{\cal G} = {\cal F}
- \int d^2r \mu(\sigma) n_0({\bf r},\sigma),
\label{12}
\end{equation}
in which $\mu(\sigma)$ is used to tune $g(\sigma)$.  One may now
freely extremalize ${\cal G}$ over all $n_0$, constrained only by
the normalization $\int d\sigma n_0({\bf r},\sigma) = 1$, to
obtain
\begin{equation}
n_0({\bf r},\sigma) = e^{-\bar \beta W(\Psi_0 - \alpha)}
e^{\bar \beta [\mu(\sigma) - \sigma (\Psi_0-\alpha)]},
\label{13}
\end{equation}
where $\Psi_0({\bf r}) = \delta {\cal H}/\delta Q_0({\bf r})$
is the equilibrium stream function, $\alpha({\bf r}) =
\alpha_0 \lambda({\bf r})$, and
\begin{equation}
W(\tau) = \frac{1}{\bar \beta} \ln\left\{\int d\sigma
e^{\bar \beta [\mu(\sigma) - \sigma \tau]} \right\}.
\label{14}
\end{equation}
Substituting (\ref{13}) back into (\ref{12}) and using (\ref{4}),
finally determines ${\cal G}$ as a functional of $\Psi_0$:
\begin{equation}
{\cal G}[\Psi_0] = {\cal L}[\Psi_0]
- \int d^2r W(\Psi_0 - \alpha).
\label{15}
\end{equation}
The latter is finally determined by the (minimum for $\bar T >
0$, maximum for $\bar T < 0$) condition
\begin{equation}
-\delta {\cal L}/\delta \Psi_0({\bf r})
\equiv Q_0({\bf r}) = F[\Psi_0({\bf r}) - \alpha({\bf r})],
\label{16}
\end{equation}
where $F(\tau) = -\partial_\tau W(\tau)$. Since $W(\tau)$ is
convex, $F(\tau)$ is monotonic.  Using (\ref{5}) and (\ref{6}),
the left hand side is $Q_0 = (-\nabla^2 + k_R^2)\Psi_0 + f$, and
(\ref{16}) represents a nonlinear PDE for $\Psi_0$. The conserved
quantities are set by the derivatives $P = -\partial {\cal
G}/\partial \alpha_0 = \int d^2r F(\Psi_0 - \alpha)$, $g(\sigma) =
-\delta {\cal G}/\delta \mu(\sigma) = \int d^2r n_0({\bf
r},\sigma)$.

Substituting (\ref{16}) into (\ref{1}), one finds $Q({\bf r},t) =
Q_0({\bf r} + {\bf \hat l} \alpha_0 t)$ or $Q_0(r,\theta +
\alpha_0 t)$, depending on the choice of $\lambda$. Fixed momentum
solutions are not in general static, but require a background flow
with velocity $\alpha_0$ along the symmetry direction.

Equations (\ref{15}), (\ref{16}) are the fundamental results of
this paper, providing a complete description of the fluid
equilibria for a rather general class of problems. Ultimately one
is interested in coherent vortex formation, where $Q$ is large in
some compact region, and is much smaller outside of it. One sees
from (\ref{3}) that these are high energy configurations, and
hence correspond to $\bar T < 0$ \cite{MWC}.

\begin{figure}[tbp]

\includegraphics[width=3.0in]{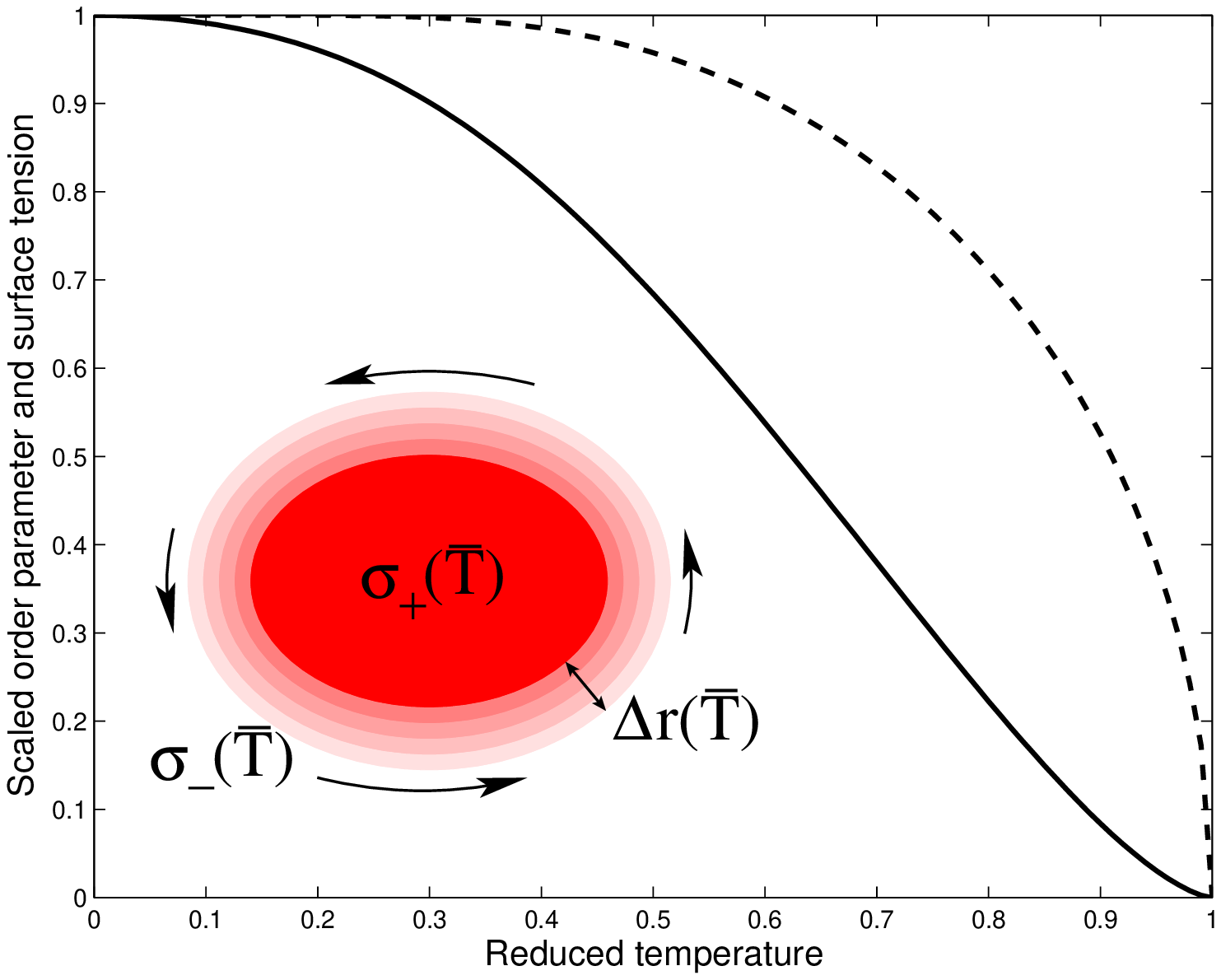}

\caption{Scaled equilibrium charge density $q_{\rm eq}(\bar t)$
(dashed line) and surface tension $\Sigma(\bar t)$ (solid line)
versus reduced temperature $\bar t = \bar T/\bar T_0$. For $\bar t
\to 0$, $q_{\rm eq} \approx 1 - 2 e^{-2/\bar t}$, $\Sigma \approx
1 - \pi^2 \bar t^2/12$, while for $\bar t \to 1$, $q_{\rm eq}
\approx \sqrt{3(1-\bar t)}$, $\Sigma \approx [2(1-\bar t)]^{3/2}$.
Inset: schematic coherent vortex with asymptotic charge density
$\sigma_\pm = \frac{1}{2} \sigma_0 (1 \pm q_{\rm eq})$ inside and
outside.  Arrows indicate the direction of fluid flow, essentially
parallel to the interface.}

\label{fig1}
\end{figure}

For tractable applications ``finite level systems'' are useful:
let $g(\sigma) = \sum_{k=1}^S A_k \delta(\sigma - \sigma_k)$, in
which $A_k$ is the total area occupied by charge $\sigma_k$. One
correspondingly requires only a finite number of Lagrange
multipliers $\mu_k$: $e^{\bar \beta \mu(\sigma)} = \sum_{k=1}^S
e^{\bar \beta \mu_k} \delta(\sigma-\sigma_k)$, and $W(\tau)$
becomes (the log of) a discrete sum. One obtains $n_0({\bf
r},\sigma) = \sum_{k=1}^S \rho_k({\bf r}) \delta(\sigma -
\sigma_k)$, $Q_0({\bf r}) = \sum_{k=1}^S \sigma_k \rho_k({\bf
r})$, where
\begin{equation}
\rho_k({\bf r}) \equiv
\frac{e^{\bar \beta \{\mu_k - \sigma_k
[\Psi_0({\bf r}) - \alpha({\bf r})]\}}}
{\sum_{l=1}^S e^{\bar \beta \{\mu_l
- \sigma_l[\Psi_0({\bf r})
- \alpha({\bf r})]\}}},
\label{17}
\end{equation}
has spatial integral $A_k$, and is therefore the equilibrium
number density for charge $\sigma_k$ \cite{foot:ptchg}.

To illustrate properties of the solutions, consider the two level
system, $\sigma_k = 0,\sigma_0$, in the beta-plane where $\alpha =
\alpha_0 y$: $\rho_2 = (e^{\bar \beta [\sigma_0(\Psi_0-\alpha)
-\mu]} + 1)^{-1}$ is the Fermi function, where $\mu = \mu_1-\mu_0$
is the chemical potential difference. Define the temperature scale
$\bar T_0 = -\sigma_0^2/4k_R^2 < 0$, and let $\bar t = \bar T/\bar
T_0$, $p_0 = -(\sigma_0 \tilde \Psi_0 - \mu)/2\bar T_0$, $q_0 =
2Q_0/\sigma_0 - 1$, ${\bm \rho} = (\rho_x,\rho_y) = k_R {\bf r}$.
In terms of these scaled variables, the free energy and
equilibrium equation take the form
\begin{eqnarray}
{\cal G} &=& -\frac{\sigma_0^2}{4k_R^4} \int d^2\rho
\left[\frac{1}{2} |\nabla_\rho p_0|^2 - h p_0 + V(p_0) +
\epsilon_0 \right]
\nonumber \\
q_0 &=& (-\nabla_\rho^2 + 1) p_0 - h = \tanh(p_0/\bar t),
\label{18}
\end{eqnarray}
where $h({\bm \rho}) = h_0 + g_0 \rho_y$, with $h_0 = 1 -
2f_0/\sigma_0 + \mu/2\bar T_0$, $g_0 = -2(\alpha_0 k_R^2 +
\beta_f)/\sigma_0 k_R$. The $p_0$-independent $\epsilon_0({\bm
\rho})$ term is unimportant. The potential $V(p_0) = \frac{1}{2}
p_0^2 - \bar t \ln[2\cosh(p_0/\bar t)]$ is an even function with a
single minimum at $p_0 = 0$ for $\bar t > 1$, and symmetric double
minima at $\pm p^0_{\rm eq}(\bar t)$ satisfying $p^0_{\rm eq} =
\tanh(p^0_{\rm eq}/\bar t) = q^0_{\rm eq}$ for $\bar t < 1$.

Equation (\ref{18}) is a standard continuum model of a binary
fluid in a \emph{gravitational field} $g_0$, composed of a mixture
of heavy ($q_0 = 1$, $Q_0 = \sigma_)$) and light ($q_0 = -1$, $Q_0
= 0$) particles. The $|\nabla p_0|^2$ term is an attractive
interaction between like particles which encourages phase
separation at low temperatures, $\bar t < 1$. For $h \neq 0$,
$V(p_0) - hp_0$ has a unique absolute minimum $p_{\rm eq}(\bar
t,h) = -p_{\rm eq}(\bar t,-h)$.  For $\bar t > 1$, $p_eq$ is
continuous through $h=0$, while for $\bar t < 1$ it jumps
discontinuously between the heavy and light phases, $\pm p_{\rm
eq}^0(\bar t)$. This analogy makes it obvious that the stable
equilibrium state must be vertically stratified, with $p_0$ and
$q_0$ increasing monotonically (lighter phases floating on denser
phases) in the direction of $g_0$ (southward if $\tilde \beta_f
\equiv \beta_f + \alpha_0 k_R^2 > 0$).  For $\bar t < 1$ there
will be a sharp interface centered at $\rho_y = -h_0/g_0$
(determined by $\mu$) of width $\Delta \rho_y \sim \bar t$ (see
below) between the segregated phases. The analogy generalizes
easily to multiple charge levels, which generate multicomponent
fluid mixtures with different combinations of separated and
unseparated phases produced as $\bar t,\mu_i$ are varied. However,
$g_0$ is unchanged, and in equilibrium the system must again be
vertically stratified with phases density ordered along $g_0$. The
argument also clearly generalizes to nonconstant $\beta_f(y)$. A
varying $g_0(y)$ still produces vertical stratification. This
establishes the claim that only equilibria with purely zonal flow
are stable in the presence of a beta effect, $\tilde \beta_f \neq
0$, and is completely consistent with inverse cascade arguments
based on conservation of $B$.

\begin{figure}[tbp]

\includegraphics[width=3.0in]{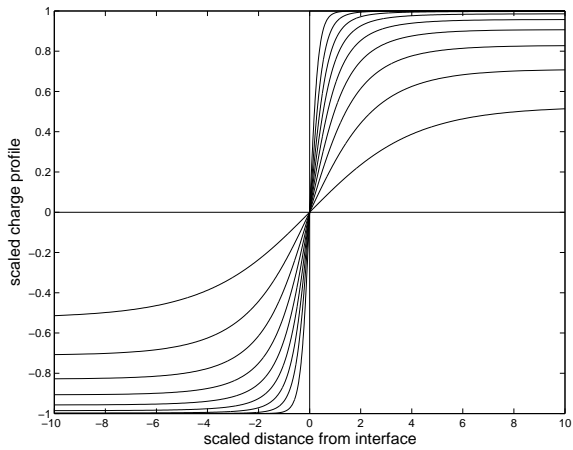}

\caption{Scaled interface profiles $q_0(\xi,\bar t)$, as a
function of the scaled normal coordinate $\xi$, for a sequence of
reduced temperatures from $\bar t = 0$ (discontinuous step, $q_0 =
{\rm sgn}(\xi)$) to $\bar t = 1$ (completely flat, $q_0 \equiv 0$)
in steps of $0.1$. For small $\bar t$, $q_0 \approx \tanh(\xi/\bar
t)$ (hence width $\Delta r \approx \bar t/k_R$), while $p_0
\approx {\rm sgn}(\xi) (1-e^{-|\xi|})$, remaining continuous even
at $\bar t = 0$. Near $\bar t = 1$, $q_0 \approx p_0 \approx
q_{\rm eq} \tanh(\xi/2\xi_0)$ with diverging width $\xi_0 = k_R
\Delta r = 1/\sqrt{2(1-\bar t)}$.}

\label{fig2}
\end{figure}

Only for $\tilde \beta_f = 0$, hence $\alpha_0 = -\beta_f/k_R^2 =
{\cal O}(10\ \rm{cm/s})$ does the gravity effect disappear
\cite{foot:gravity}. In a background flow moving at speed
$\alpha_0$, an effective isotropic $f$-plane is restored. Consider
a large vortex region whose boundary is smooth on the scale of its
width $\Delta r(\bar T)$ (see below), viewed here as a bubble of
one phase inside the other (see inset to Fig.\ \ref{fig1}).
Coexistence requires $\bar T < \bar T_0$, and $h = 0$, i.e., $\mu
= \sigma_0(\sigma_0 - 2f_0)/2k_R^2$.  Near the interface,
(\ref{18}) becomes a 1D equation in the coordinate $\xi = {\bm
\rho} \cdot {\bf \hat n}$ normal to the interface, with $q_0,p_0
\to \pm q^0_{\rm eq}$ [hence charges $\sigma_\pm(\bar T) =
\frac{1}{2}\sigma_0 (1 \pm q^0_{\rm eq})$], deep on either side.

The free energy increment per unit length $L$ of the interface,
$\Delta {\cal G}/L = (2|\bar T_0|^{3/2}/\sigma_0) \Sigma$ yields
the scaled the surface tension
\begin{equation}
\Sigma(\bar t) = \int_{-\infty}^\infty d\xi (\partial_\xi p_0)^2.
\label{20}
\end{equation}
In Fig.\ \ref{fig1} numerical solutions for $q^0_{\rm eq}$,
$\Sigma$ are plotted (and exact asymptotics described), while in
Fig.\ \ref{fig2} scaled interface profiles $q_0(\xi)$ are plotted
for several $\bar t$. The true equilibrium solution is finally
obtained by minimizing the vortex perimeter $L$ at fixed area $A$
(set by the total charge $\Omega_1$), yielding, not surprisingly,
a circular vortex.  These arguments again generalize to multiple
charge levels, where one could in principle have more than two
phases in simultaneous equilibrium, with vortices composed of a
central core of one phase, ringed by one or more other phases.

Since $B$ is no longer conserved in the $f$-plane, there is no
dynamical barrier to the formation of isotropic structures, again
establishing consistency between arguments based on the dynamics
of the inverse cascade, and those based on purely equilibrium
thermodynamics.  It appears likely that conservation of $B$ is a
microscopic reflection of the analogy to gravity-induced density
stratification.  Since vortices drift across $g_0$, rather than
accelerating along it, equilibration is necessarily far less
direct than in physical binary fluid systems, and $B$ may be one
of the mechanisms controlling this. This idea will be investigated
more carefully in future work.


The author thanks A. M. Balk and R. E. Glazman for highly
informative discussions, and gratefully acknowledges the
hospitality of the Aspen Center for Physics where this work was
initiated.


\begin{thebibliography}{99}
\bibitem{foot:sym} Formally, one chooses $\alpha$ in such a way that
$\{Q,L\} = \partial_\xi Q$, where $\xi$ is the symmetry
coordinate, so $L$ is the generator of translations along $\xi$.
\bibitem{foot:gravity} Since $\beta_f(y) \equiv f'(y)$ decreases
with increasing $|y|$, $f(y) - \beta_f(y_0)(y-y_0) \approx
f(y_0)-\frac{1}{2} f''(y_0)(y-y_0)^2$ is not perfectly flat, but
has an extrememum.  In the northern (southern) hemisphere it traps
lighter (denser) phases, while denser (lighter) phases will be
pushed to higher and lower latitudes. Only lower (higher) charged
coherent vortices may then be truly stable.
\end{thebibliography}
\end{document}